\documentstyle[aps,prl,epsfig]{revtex}

\begin{document}
\twocolumn[\hsize\textwidth\columnwidth\hsize\csname@twocolumnfalse\endcsname
\draft

\title{Long-Range Coulomb Effect on the Antiferromagnetism in Electron-doped Cuprates}

\author{Xin-Zhong Yan}

\address{Institute of Physics, Chinese Academy of Sciences, P.O. Box603, Beijing 100080, China\\}
%E-mail: yanxz@aphy.iphy.ac.cn

\date{\today}

\maketitle

\widetext
\begin{abstract}
Using mean-field theory, we illustrate the long-range Coulomb effect on the antiferromagnetism in the electron-doped cuprates. Because of the Coulomb exchange effect, the magnitude of the effective next nearest neighbor hopping parameter increases appreciably with increasing the electron doping concentration, raising the frustration to the antiferromagnetic ordering. The Fermi surface evolution in the electron-doped cuprate Nd$_{2-x}$Ce$_x$CuO$_4$ and the doping dependence of the onset temperature of the antiferromagnetic pseudogap can be reasonably explained by the present consideration.

\end{abstract}


\vfill
\narrowtext

\vskip2pc] 

The Fermi surface (FS) evolution with electron doping in Nd$_{2-x}$Ce$_x$CuO$_4$ has been observed by angle-resolved photoemission spectroscopy (ARPES) experiments\cite{Armitage,Damascelli}. Recently, much attention has been paid to understanding the physics of this phenomenon\cite{Kusko,Markiewicz,Kusunose,Kyung,Senechal,Yuan}. At low temperatures, the electron-doped cuprates are in the states of antiferromagnetic (AF) phase within a wide doping region\cite{Luke,Onose}. The FS evolution is therefore closely relevant to the antiferromagnetic correlation. From the framework of the Hubbard model, the doped electrons occupy the upper Hubbard band. On the other hand, according to the experimental observation, with increasing electron doping, the energy gap should decrease and eventually close up at the optimal doping, $x \approx 0.14$, where the AF phase terminates. Therefore, to interpret the experimental results, one needs to assume the on-site interaction $U$ in the Hubbard model to be dramatically decreasing with increasing electron doping concentration. This is a puzzle within the Hubbard model with constant on-site interaction. Some investigators have treated the doping dependence of $U$ by considering some kind of screening\cite{Kusko,Markiewicz,Kyung}. 

In this work, we explore the long-range Coulomb effect (LRCE) on the AF ordering. Due to the Coulomb exchange effect, the LRCE results in excess electron hopping. As a whole, the excess hopping tends to frustrate the AF ordering. With increasing the electron doping, this exchange effect becomes significant, leading to the decreasing of the energy gap. We will see this gives a reasonable explanation of the FS evolution and the envelope of the onset temperature of the antiferromagnetic pseudogap in electron-doped Nd$_{2-x}$Ce$_x$CuO$_4$.

We start with the following two-dimensional square lattice model
\begin{equation}
H= - \sum_{ij,\sigma}t_{ij} c_{i\sigma}^{\dagger}c_{ j\sigma} 
+ U \sum_{i}n_{i\uparrow} n_{i\downarrow}
+ {1\over 2}\sum_{i\not=j}v_{ij}n_{i} n_{j}
\label{Hamiltonian}
\end{equation}
where $t_{ij}$ denotes the hopping energy of an electron between the lattice sites $i$ and $j$, $c_{i\sigma}^\dagger$ ($c_{i\sigma}$) represents the electron creation (annihilation) operator of spin-$\sigma (= \pm 1$ for up and down spins, respectively) at site $i$, $n_{i\sigma}=c_{i\sigma}^\dagger c_{i\sigma}$, and $n_{i}=n_{i\uparrow}+n_{i\downarrow}$ is the electron density operator. Besides the on-site Hubbard repulsion $U$, we take into account the long-range Coulomb interaction in the third term. Such a similar type of Hamiltonian has been used to investigate the LRCE on the $d$-wave pairing for the hole-doped case\cite{Esirgen}. 

In the hopping term, besides the nearest neighbor (n.n) hopping, the other hopping processes within a range need to be included. The essential role of the next n.n hopping for the validity of the single-band Hubbard model for describing the cuprates has been investigated by comparing it with the two-band\cite{Macridin} and three-band Hubbard models\cite{Markiewicz}. The particle-hole asymmetry in the cuprates can be understood by taking into account the next n.n hopping \cite{TM,Gooding}. By including the next and third n.n hopping parameters in the types of $t-J$ and Hubbard models, numerous studies have been carried out to explore the properties of the electron-doped cuprates\cite{Kusko,Markiewicz,Kusunose,Kyung,Senechal,Yuan,Tohyama,Lee,LZL,Yuan1}. We here take two more additional parameters for the fourth and fifth hopping, each of which is smaller than the formers. In terms of the n.n hopping parameter, $t_{(1,0)} \equiv t$, the values of other 4 parameters are given as $t_{(1,1)} = -0.325t$, $t_{(2,0)} = 0.17t$, $t_{(2,1)} = -0.121t$, and $t_{(2,2)} = -0.07t$, where the subscripts instead of $ij$ denote the components of a vector of length $|{\bf i} - {\bf j}|$. The values of the next and third n.n hopping parameters are approximately the same as that in the literatures\cite{Kusko,Markiewicz,Kusunose,Kyung,Senechal,Yuan,Tohyama,Lee,LZL,Yuan1}. From the hopping term, the single-particle dispersion $\epsilon^0_k$ is given by
$$
\begin{array}{rl}
\epsilon^0_k =& -2t(c_x+c_y)-4t_{(1,1)}c_xc_y-2t_{(2,0)}(c_{2x}+c_{2y})\cr\cr
& -4t_{(2,1)}(c_{2x}c_y+c_{2y}c_x)-4t_{(2,2)}c_{2x}c_{2y}
\end{array}
$$
with $c_{lx}=\cos(lk_x)$. 

For the long-range Coulomb interaction $v_{ij}$, we take the following form
\begin{equation}
v_{ij}=V_1\exp(-|{\bf i}-{\bf j}|/d)/|{\bf i}-{\bf j}|,
\label{LRCE}
\end{equation}
where the length scale is in unit of the lattice constant. Here we set $V_1 = 1.0t$ and $d$ = 4 for the present calculation. For the on-site $U$, we use $U = 4.8t$, which is within the range of the doping-dependent interaction adopted in the existing calculations\cite{Kusko,Markiewicz,Kyung}. The strength of $V_1$ with $V_1/U \approx 0.21$ is a typical one\cite{Esirgen}. 

By the mean-field (MF) approximation, from the long-range Coulomb term, we get the exchange part of the self-energy as
\begin{equation}
\Sigma^x_k = -\frac{1}{N}\sum_{k'}v(|{\bf k}-{\bf k'}|)\langle c_{k'\uparrow}^{\dagger}c_{k'\uparrow}\rangle 
\label{xc}
\end{equation}
where $N$ is the number of lattice sites, $v(|{\bf k}-{\bf k'}|)$ is the Fourier transform of $v_{ij}$, and $\langle\cdots\rangle$ denotes an average. This exchange self-energy is equivalent to a hopping energy in real space. The corresponding hopping integral is given as $t^x_{ij} = v_{ij}\langle c_{i\uparrow}^{\dagger}c_{j\uparrow}\rangle$. The significance of this exchange effect and its doping dependence will be discussed later. With the contribution from the exchange self-energy, the effective single-particle dispersion is given by $\epsilon_k = \epsilon^0_k + \Sigma^x_k$.
 
At low temperatures, the strong on-site Coulomb interaction leads to the AF ordering. In the case of the electron doping under consideration, the AF ordering is a commensurate spin-density wave. The order parameter is given by $\Delta = U\langle n_{\downarrow}(Q)-n_{\uparrow}(Q)\rangle /2N$, where $n_{\sigma}(Q)$ is the Fourier transform of the electron density of spin-$\sigma$ at wave vector $Q = (\pi,\pi)$. By the MF approximation, the energy spectrum of the quasiparticle is then given by
\begin{equation}
E^{\pm}_k =(\epsilon_k+\epsilon_{k+Q}\pm E_k)/2,
\label{ESP}
\end{equation}
where the + (-) sign refers to the upper (lower) Hubbard band, and $E_k = \sqrt{(\epsilon_k-\epsilon_{k+Q})^2+4\Delta^2}$. The parameter $\Delta$ and the chemical potential $\mu$ are self-consistently determined by 
\begin{eqnarray} 
\frac{U}{N}\sum_{k}[f(E^{-}_k)-f(E^{+}_k)]/E(k)=1,\cr\cr
\frac{1}{N}\sum_{k}[f(E^{-}_k)+f(E^{+}_k)]=n,
\end{eqnarray}
where $f$ is the Fermi distribution function, and $n$ is the electron density. In terms of doping concentration $x$, $n = 1 + x$. 

To analyze the ARPES observations on the FS evolution, one usually considers the spectral density occupied by the electrons. By the MF theory, this spectral density is given by
\begin{equation}
A^{<}(k,\omega) = f(\omega)[u^2(k)\delta(\omega-E^{+}_k)+v^2(k)\delta(\omega-E^{-}_k)]
\end{equation}
with $u^2(k)=1-v^2(k)=[1+(\epsilon_k-\epsilon_{k+Q})/E_k]/2$. Fig. 1 shows the Fermi-level electron distributions in momentum space obtained by integrating $A^{<}(k,\omega)$ in a energy window $|\omega-\mu| < 0.15t$ at temperature $T = 0.1t$. At low doping, the FS (in the first quadrant of the Brillouin zone) appears as two small pockets centered at $(\pi,0)$ and $(0,\pi)$. With increasing doping, these pockets extend to larger squares. Such an evolution of the pockets reflects the occupation of the doped electrons at the upper Hubbard band. On the other hand, a small pocket begins to form around $(\pi/2,\pi/2)$ at $x \sim 0.1$ and grows up with increasing doping. The formation of this pocket stems from the contribution of the lower Hubbard band; with increasing doping, the energy gap decreases and meanwhile the lower Hubbard band shifts up toward to the FS. At high doping, the energy gap closes up, therefore the FS is a single curve centered at $(\pi,\pi)$ ($x$ = 0.18). Clearly, the present calculation reproduces the results of the existing calculations based on the doping-dependent $U$ Hubbard model\cite{Kusko,Kyung} and is in agreement with the experimental observations\cite{Armitage}. By the $t-t'-t''-U$ model, a constant $U$ actually results in a nearly constant energy gap in a wide region of the doping concentration, and hence cannot explain the FS evolution. Though $U$ is constant in the present calculation, but because of the LRCE, the energy gap decreases remarkably with increasing doping. We will discuss this problem later again. 

%\vskip -2mm
\begin{figure}
\centerline{\epsfig{file=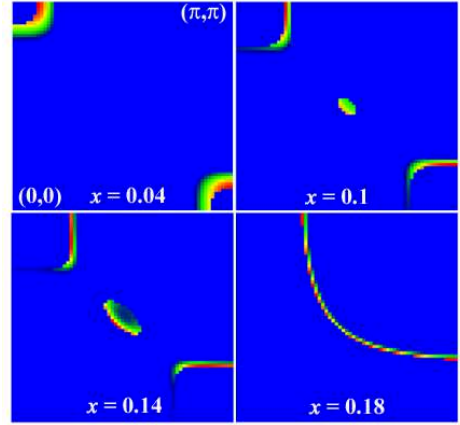,width=6. cm}}
\caption{(Color online) Maps of Fermi surface at various doping concentrations at temperature $T = 0.1t$. Highs are denoted by red and lows by blue.}
\end{figure}

In Fig. 2, we depict the MF N\'eel temperature $T_N$ (solid line with circles) as a function of the doping concentration $x$, and compare it with the result of the extended Hubbard model (EHM, in which the LRCE is not included) and the experimental data. It is seen that the difference between the present calculation with LRCE and the EHM is remarkable. By the EHM, the dependence of $T_N$ on $x$ is very weak in the region $0 < x < 0.2$. In contrast to the EHM, $T_N$ given by the present calculation with LRCE decreases distinctly with increasing doping. In particular, it drops sharply at $x \approx 0.141$. This doping concentration corresponds to the quantum critical point at which the zero-temperature AF transition terminates and is consistent with the experimental observation\cite{Luke,Onose}. Also, the envelope of $T_N$ with LRCE is in fairly good agreement with the experimental results for the pseudogap $\Delta_{\rm pg}$ and its onset temperature $T^*$\cite{Onose}. The MF N\'eel transition is associated with the appearance of the local magnetization, and hence the comparison between the MF $T_N$ and $T^*$ or $\Delta_{\rm pg}$ is meaningful\cite{Markiewicz}. The MF order parameter is a measure of the local order of the real system with long-wavelength fluctuation, similar to the case of superconducting ordering\cite{Chen,Yan}. 

\vskip -4mm
\begin{figure}
\centerline{\epsfig{file=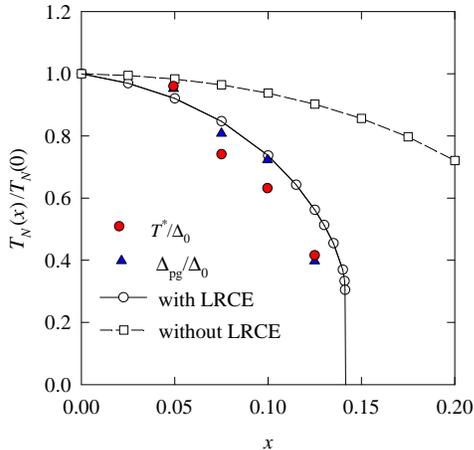,width=8. cm}}
\vskip -6mm
\caption{(Color online) Normalized MF N\'eel temperature $T_N(x)/T_N(0)$ as function of the doping concentration $x$. The solid line with circles is the present calculation taking into account the LRCE. The dashed line with squares is the result of the extended Hubbard model without the long-range Coulomb interaction. The pseudogap $\Delta_{\rm pg}$ and its onset temperature $T^*$ (both of them normalized by $\Delta_0 = 460 K$) are the experimental results [10].}
\end{figure}

Why does the LRCE lead to $T_N$ decreasing with increasing doping? To answer this question, we here consider the excess hopping parameter $t^x_{ij}$ due to the Coulomb exchange. This quantity can be written as 
\begin{equation}
t^{x}_{ij} \equiv t^x_l = {v_{ij}\over 2N}\sum_k\langle c^{\dagger}_{k\uparrow}c_{k\uparrow}\rangle S_l(k)
\label{xchop}
\end{equation}
where $S_l(k) = \cos(l_xk_x)\cos(l_yk_y)+\cos(l_yk_x)\cos(l_xk_y)$ and again $l = (l_x,l_y)$ represents a vector of length $|{\bf i}-{\bf j}|$. The doping dependence of $t^x_{ij}$ originates from the Fermi distribution $\langle c^{\dagger}_{k\uparrow}c_{k\uparrow}\rangle$. At low temperature, the contribution to the integral in Eq. (\ref{xchop}) comes from the Fermi area. The Fermi surface varies with changing doping concentration. Especially, with increasing doping, the Fermi area increases considerably in the regions near the points $(\pi,0)$ and $(0,\pi)$ since at which the upper Hubbard band reaches its minimum and where the energy dispersion is nearly flat. Therefore, the variation of $t^x_l$ stems predominantly from the integral in these regions. On the other hand, with the change of the Fermi surface, the factor $S_l(k)$ varies slowly for small $l$, but rapidly for large $l$. It is then easy to see the variation of $t^x_l$ with increasing doping at low temperature: for example, $t^x_{(1,0)}$ is almost unchanged for $S_{(1,0)}(\pi,0)=0$; $t^x_{(1,1)}$ has a large variation because of $S_{(1,1)}(\pi,0)=-2$. Since the original hopping parameter $t_{(1,1)}$ is negative, the magnitude of the effective hopping parameter $t_{(1,1)}+t^x_{(1,1)}$ should be enhanced with increasing doping. For large $l$, because of the cancellation from the destructive factor $S_l(k)$, not only the variation, but also the magnitude of $t^x_l$ are negligible small. At high temperature, since the integral in Eq. (\ref{xchop}) is taken over a spreading area wider than the Fermi area, the behavior of $t^x_l$ is not so intuitive. Fig. 3 exhibits the variations of the magnitudes of the effective hopping parameters $t_l +t^x_l$ as functions of the doping concentration $x$ at temperature $T = 0.53t$. The quantity $\delta t_l(x)$ is defined as  
\begin{equation}
\delta t_l(x) = [t^x_l(x)-t^x_l(0)]{\rm sgn}[t_l+t^x_l(0)]. 
\end{equation}
From Fig. 3, it is seen that except for $\delta t_{(1,0)}$ that is nearly constant, the other parameters vary approximately linearly with $x$. Especially, among those effective hopping parameters only the magnitude of $t_{(1,1)}+t^x_{(1,1)}\equiv t'$ increases appreciably with increasing $x$. This is consistent with the above analysis. It is know that a large ratio $t'/t$ can destroy the AF instability at weak $U$\cite{Hofstetter}. This is also true for strong on-site interactions. For large $t'/U$, the next n.n. AF coupling constant is $J' = 4 t'^2/U$, which leads to the frustration of AF ordering in the square lattice. It is therefore clear that the AF order parameter and the transition temperature decrease with increasing doping concentration. 

\vskip -3mm
\begin{figure}
\centerline{\epsfig{file=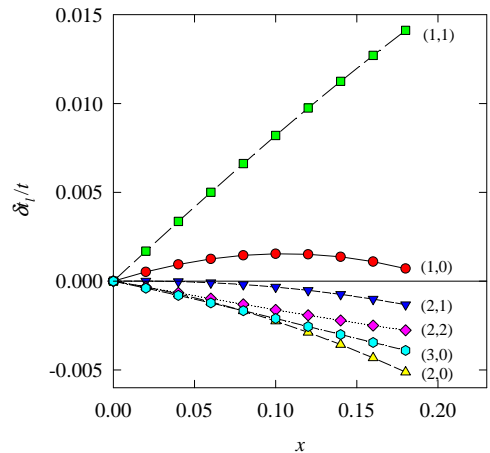,width=8. cm}}
\vskip -6mm
\caption{(Color online) Variation of the effective hopping parameter $\delta t_l$ as function of the doping concentration $x$ at temperature $T = 0.53t$. A pair of numbers with brackets denotes the vector $l = (l_x,l_y)$.}
\end{figure}

Shown in Fig. 4 is the parameter $t^x_l$ as a function of distance at the MF $T_N = 0.39t$ at $x = 0.1$. It is seen that only the magnitudes of the first 4 parameters are appreciable; from the fifth to the one at distance 4, all of them are very small; after that all other parameters are negligible small. In a wide region on the temperature-doping concentration phase diagram, the parameter $t^x_l$ behaves almost the same as shown in Fig. 4, with only a visual change in the next n.n hopping parameter. From the behavior of the parameter $t^x_l$, the range of the predominant Coulomb exchange effect seems to be shorter than 3. Of course, these results are obtained for the particular sets of the input parameters ($t_l$'s, $U$, $V_1$ and $d$). But these parameters are reasonable. In the present calculation, all the input parameters were not fine-tuned. 

\vskip -3mm
\begin{figure}
\centerline{\epsfig{file=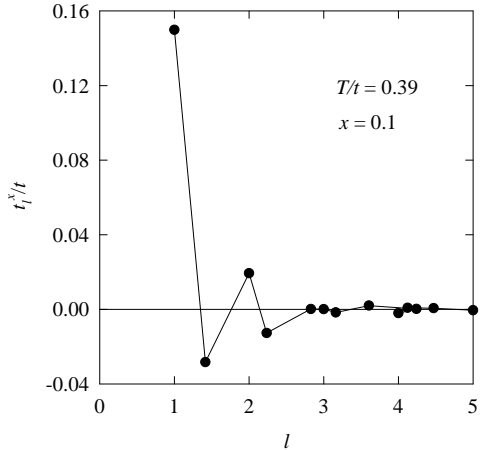,width=8. cm}}
\vskip -6mm
\caption{The hopping parameter $t^x_{l}$ due to Coulomb exchange as function of the distance $l$ at $T = 0.39t$ and $x = 0.1$.}
\end{figure}

In summary, we have investigated the long-range Coulomb effect on the antiferromagnetism in the electron-doped cuprates using the mean-field theory. Due to the Coulomb exchange, the magnitude of the effective next nearest neighbor hopping parameter in especial increases appreciably with increasing the electron doping concentration. This leads to stronger frustration to the AF ordering at higher doping concentration. Therefore the transition temperature decreases with increasing doping. Consequently, the AF phase terminates at a doping concentration $x \approx 0.14$ in consistent with experiments. The present calculation gives a reasonable explanation to the doping dependence of the onset temperature of the AF pseudogap as well as to the FS evolution in the electron-doped cuprate Nd$_{2-x}$Ce$_x$CuO$_4$.

The author thanks Prof. C. S. Ting for useful discussions on the related problems. This work was supported by Natural Science Foundation of China under grant number 10174092 and by Department of Science and Technology of China under grant number G1999064509.

%\vskip 4mm

\end{document}